\documentclass[%
 reprint,
 superscriptaddress,
 nofootinbib,
 twocolumn, 
 amsmath,amssymb,
 aps,
prb,
]{revtex4-2}

\usepackage[colorlinks = true,
            linkcolor = blue,
            urlcolor  = red,
            citecolor = red,
            anchorcolor = blue]{hyperref}
\usepackage[T1]{fontenc}
\usepackage[utf8]{inputenc}
\usepackage{todonotes}
\usepackage{graphicx}% Include figure files
\usepackage{dcolumn}% Align table columns on decimal point
\usepackage{color}
\usepackage{amsmath}
\usepackage{braket}
\usepackage{appendix}
\usepackage{MnSymbol}
\usepackage{orcidlink}
\usepackage{glossaries}
\newacronym{CDW}{CDW}{charge-density-wave}
\newacronym{RIXS}{RIXS}{resonant inelastic x-ray scattering}
\newacronym{DMRG}{DMRG}{density matrix renormalization group}
\newacronym{DQMC}{DQMC}{determinant quantum Monte Carlo}
\newacronym{QMC}{QMC}{quantum Monte Carlo}
\newacronym{SSH}{SSH}{Su-Schrieffer-Heeger}
\newacronym{oSSH}{oSSH}{optical Su-Schrieffer-Heeger}
\newacronym{2D}{2D}{two-dimensional}
\newacronym{1D}{1D}{one-dimensional}
\newacronym{FS}{FS}{Fermi surface}
\newacronym{eph}{$e$-ph}{electron-phonon}
\newacronym{HMC}{HMC}{hybrid Monte Carlo}
\newacronym{KVB}{KVB}{Kekul{\'e} Valence Bond}
\newacronym{KVBS}{KVBS}{Kekul{\'e} Valence Bond Solid}
\newacronym{GS}{GS}{ground state}
\newacronym{QCP}{QCP}{quantum critical point}
\newacronym{FSS}{FSS}{finite-size scaling}
\newacronym{AFM}{AFM}{antiferromagentic}
\newacronym{SC}{SC}{superconducting}
\newacronym{HSSH}{HSSH}{Hubbard-SSH}
\newacronym{SM}{SM}{supplementary materials}
\newacronym{HH}{HH}{Hubbard-Holstein}
\newacronym{BE}{BE}{binding energy}
\newacronym{BOW}{BOW}{bond ordered wave}
\newacronym{INS}{INS}{inelastic neutron scattering}
\newacronym{e-e}{$e$-$e$}{electron-electron}

\begin{document}

\preprint{}
\title{Enhanced carrier binding and bond correlations in the {H}ubbard-{S}u-{S}chrieffer-{H}eeger model with dispersive optical phonons}

\author{Debshikha Banerjee\orcidlink{0009-0001-2925-9724}}
\affiliation{Department of Physics and Astronomy, The University of Tennessee, Knoxville, Tennessee 37996, USA}
\affiliation{Institute for Advanced Materials and Manufacturing, University of Tennessee, Knoxville, Tennessee 37996, USA\looseness=-1}

\author{Alberto Nocera\orcidlink{0000-0001-9722-6388}}
\affiliation{Department of Physics Astronomy, University of British Columbia, Vancouver, British Columbia, Canada V6T 1Z1}
\affiliation{Stewart Blusson Quantum Matter Institute, University of British Columbia, Vancouver, British Columbia, Canada
V6T 1Z4}

\author{Steven Johnston\orcidlink{0000-0002-2343-0113}}
\affiliation{Department of Physics and Astronomy, The University of Tennessee, Knoxville, Tennessee 37996, USA}
\affiliation{Institute for Advanced Materials and Manufacturing, University of Tennessee, Knoxville, Tennessee 37996, USA\looseness=-1}

\date{\today}% It is always \today, today,
             %  but any date may be explicitly specified

\begin{abstract}
\Gls*{eph} interactions play a crucial role in determining many properties of materials. In this context, the \gls*{SSH} model, where atomic motion modulates the electronic hopping, has gained significant attention due to its potential for strong electron pairing in relation to high-$T_\mathrm{c}$ superconductivity. Previous studies of the \gls*{SSH} models have addressed many aspects of this problem, but have focused heavily on either dilute or half-filled models with dispersionless (Einstein) phonons. Here, we study the effects of dispersive optical phonons on the lightly doped one-dimensional optical Hubbard-\gls*{SSH} model using the density matrix renormalization group. We observe a significant enhancement in singlet binding driven by phonon dispersion; however, by calculating various correlation functions, we find that the enhanced binding does not translate to increased superconducting correlations but rather robust bond correlations in the studied parameter regime. Nevertheless, the significant impact of phonon dispersion on these correlations highlights the need to go beyond the Einstein phonon limit while modeling realistic quantum materials.
\end{abstract}

\glsresetall
\maketitle
 
%%%%%%%%%%%%%%%%%%%%%%%%%%%%%%%%%%%%%%%%%%%%%%%%%%%%%%%%%%%
% INTRODUCTION
%%%%%%%%%%%%%%%%%%%%%%%%%%%%%%%%%%%%%%%%%%%%%%%%%%%%%%%%%%%
\noindent{\bf Introduction} --- 
\Gls*{eph} coupling is one of the most fundamental interactions in condensed matter physics and plays a key role in conventional superconductivity~\cite{bardeen1957microscopic,bardeen1957theory}. Two widely studied \gls*{eph} models are the Holstein~\cite{Holstein2000706, HOLSTEIN1959343} and \gls*{SSH}~\cite{SSH_originalpaper_PRL_1979} models. The \gls*{SSH} type of interaction, where atomic motion modulates the electronic hopping, has gained significant attention recently due to theory studies suggesting that it can support the formation of mobile (bi)polarons~\cite{marchand2010sharp,sous2018light} in the dilute limit and could lead to high-$T_\text{c}$ superconductivity~\cite{zhang2023bipolaronic}. Beyond dilute fillings, the interplay between \gls*{e-e} and \gls*{eph} interactions can lead to several novel phases. For example, the half-filled \gls*{2D} \gls*{HSSH} model has competing \gls*{AFM} and \gls*{BOW} phases~\cite{Cai2021antiferromagnetism,Cai2022robustness,Feng2022phase,gotz2022valence,tanjaroon2025antiferromagnetic,MalkarugeCosta2024Kukule}. \gls*{SSH}-like interactions can also drive novel \gls*{CDW}~\cite{li2020quantum,cohen2023hybrid} and quantum spin liquid phases~\cite{cai2025quantum}.

Most studies of \gls*{SSH} models have considered dispersionless Einstein phonons. While this simplification is sometimes driven by the increased computational complexity associated with treating dispersive modes, it is also believed to be a good approximation for systems where the phonon bandwidth is small compared to the average frequency~\cite{de1984numerical}. However, optical phonons can have a large bandwidth in real materials, often reaching a significant fraction of the average phonon energy~\cite{Mohr2007phonon, Pintschovius, Graf2008bond, Wehinger2016soft}. 
At the same time, studies of dispersive models in the dilute limit have found that comparable dispersions can alter model predictions~\cite{marchand2013effect,costa2018phonon,bonvca2021dynamic,bonvca2022electron,kovavc2024light,kovavc2025spectral, zhang2023effect,zhang2025effects}. For example, in the Hubbard-Holstein model~\cite{kovavc2024light,kovavc2025spectral}, a dispersive phonon branch with low-energy modes at the zone center significantly enhances the formation of nearest-neighbor bipolarons, whereas low-energy dispersive modes at the zone boundary reduce the polaron's effective mass in the bond-\gls*{SSH} model~\cite{zhang2023effect}. Light (bi)polarons have also been reported for the nondispersive bond \gls*{HSSH} model in the dilute electron density limit~\cite{sous2018light, marchand2010sharp}. Given these findings, it is natural to ask whether similar strong binding persists at higher carrier concentrations and how it is affected by the phonon dispersion. 

Here we investigate dispersive optical phonons in a \gls*{1D} doped \gls*{HSSH} model using \gls*{DMRG}. We primarily focus on the lightly hole-doped regime ($\langle n \rangle = 1-\rho$, $\rho \ll 1$) relevant to many transition metal oxides. (For example, our results are directly applicable to the recently synthesized \gls*{1D} cuprate chain compounds~\cite{Chen2021anomalously}.) As with the dilute limit cases, we find that a sizable phonon dispersion has a substantial effect on the system's ground and excited state properties. Specifically, we observe that a dispersion that softens modes at $q \approx 2k_\mathrm{F}$ enhances electron binding. Notably, our calculated binding energies are larger than those obtained in the extended Hubbard model with an attractive nearest-neighbor interaction, which was recently proposed as a low-energy effective model for \gls*{1D} cuprate chains~\cite{Chen2021anomalously} and ladders \cite{padma2025beyond, scheie2025cooper}. The enhanced binding occurs in the singlet channel and produces a spin gap in the system; however, upon further analysis, we find that the binding does not lead to enhanced superconducting correlations. \\

\begin{figure}
    \centering
    \includegraphics[width=\columnwidth]{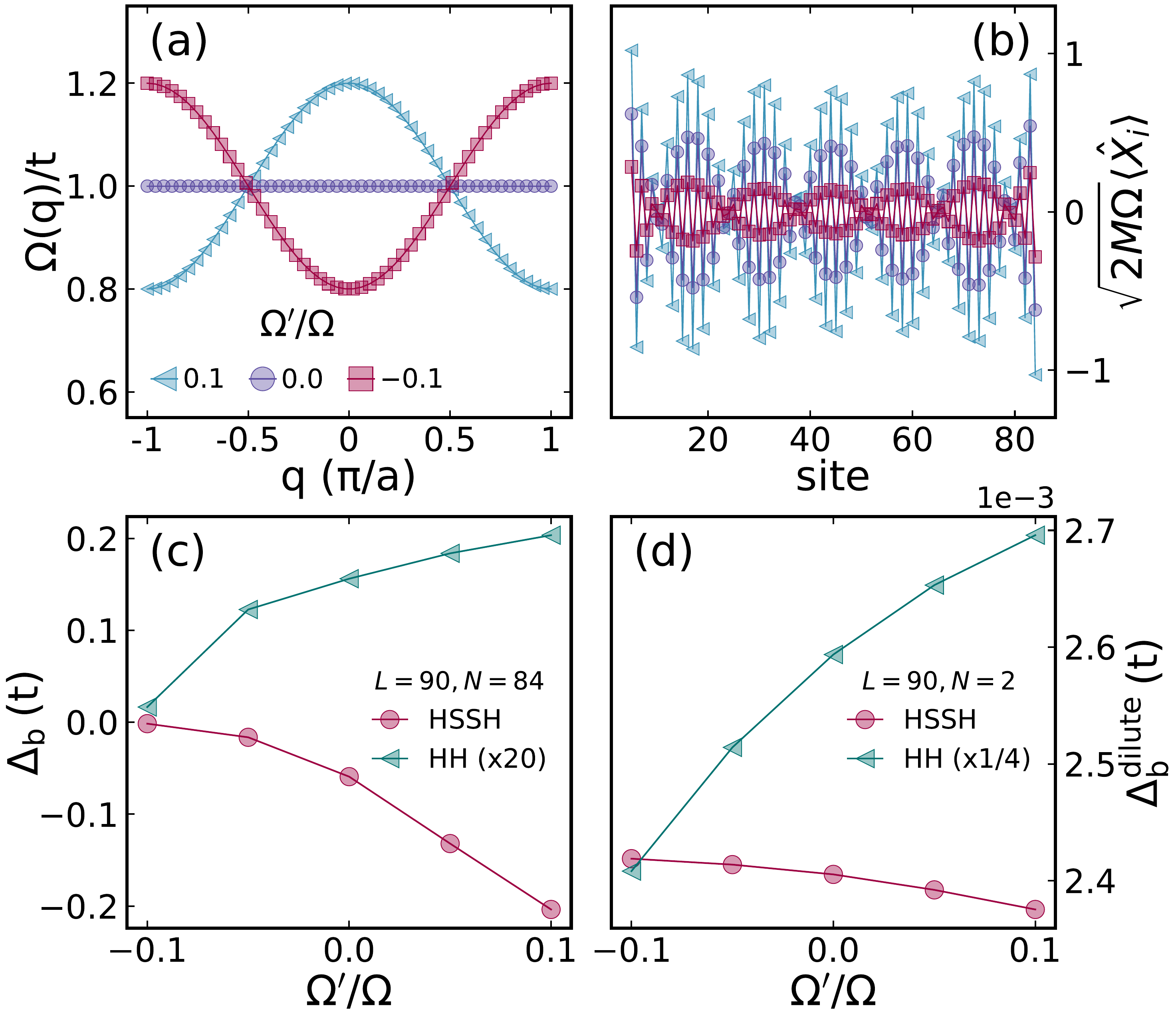}
    \caption{(a) Phonon dispersion of the optical branch with different values of $\Omega^{\prime}/\Omega$. (b) Ground state expectation values of the lattice distortion for the doped HSSH model for the same values of $\Omega^{\prime}/\Omega$ as in panel (a). (c-d) Binding energy $\mathrm{\Delta_{\mathrm{b}}}$ as a function of $\Omega^{\prime}/\Omega$ for the \gls*{HSSH} and \gls*{HH} models. All results were obtained for an $L=90$ site 1D chain with $U=8t$, $\Omega=t$, $g=0.4$ (HSSH), $g=1.13$ (HH), (b-c) $\rho = 6.67\%$, and (d) $N=2$, with different values of $\Omega^{\prime}/\Omega$ as indicated.}
    \label{fig:bare_phonon_dispersion}
\end{figure}

%%%%%%%%%%%%%%%%%%%%%%%%%%%%%%%%%%%%%%%%%%%%%%%%%%%%%%%%%%%
% MODEL and Methods
%%%%%%%%%%%%%%%%%%%%%%%%%%%%%%%%%%%%%%%%%%%%%%%%%%%%%%%%%%%
\noindent{\bf Model \& Methods} --- We study the optical variant of the \gls*{1D} \gls*{HSSH} Hamiltonian with dispersive phonons ~\cite{SSH_originalpaper_PRL_1979,Capone1997small}. It is defined as $\hat{H} = \hat{H}_e + \hat{H}_\text{ph} + \hat{H}_{e-\text{ph}}$, where  
\begin{align}\nonumber
    \hat{H}_e&=-t \sum_{i,\sigma} \left[
    c^\dagger_{i,\sigma}c^{\phantom\dagger}_{i+1,\sigma} + \text{H.c.}\right] + U \sum_{i}\hat{n}_{i,\uparrow}\hat{n}_{i,\downarrow}, \\\nonumber
    \hat{H}_\text{ph}&= \hbar\Omega \sum_{i}(b^\dagger_i b^{\phantom\dagger}_i+\tfrac{1}{2})  + \hbar\Omega^{\prime}\sum_i (b_i^{\dagger}b_{i+1}^{\phantom\dagger} + \text{H.c.}),~\text{and}\\
   \hat{H}_{e-\text{ph}}&= g\sum_{i,\sigma} \left[c^\dagger_{i,\sigma}c^{\phantom\dagger}_{i+1,\sigma}\left(\hat{X}_i - \hat{X}_{i+1}\right) + \text{H.c.}\right].
   \label{eq:Hamiltonian_HSSH}
\end{align}
Here, $c^\dagger_{i,\sigma}$ ($c^{\phantom\dagger}_{i,\sigma}$) creates (annihilates) a spin-$\sigma$ electron on lattice site $i$, $b^\dagger_i$ ($b^{\phantom\dagger}_{i}$) creates (annihilates) a local phonon quanta at site $i$, $\hbar\Omega$ is the characteristic energy of the local oscillator, 
$\hat{X}_i = \sqrt{\frac{\hbar}{2M\Omega}}(b^\dagger_i + b^{\phantom\dagger}_i)$ is the displacement operator for the $i$\textsuperscript{th} atom, $\hbar\Omega^{\prime}$ controls phonon hopping processes (and thus branch's bandwidth), $\hat{n}_i = \sum_{\sigma} \hat{n}_{i,\sigma}$ is the local electron density operator, $t$ is the nearest-neighbor electron hopping, $U$ is the on-site Hubbard repulsion, and $g$ is the \gls*{eph} coupling strength. The optical phonon branch has a dispersion $\Omega(q) = \Omega + 2\Omega^{\prime}\cos(qa)$, where $a$ is the lattice constant. The dispersion is sketched in Fig.~\ref{fig:bare_phonon_dispersion}(a) for different values of $\Omega^{\prime}$. For $\Omega^\prime > 0$ ($<0$), the phonon branch acquires significant dispersion with soft (hard) modes at $q=\pi/a$. Throughout this work, we set $\hbar = M = t = a=1$ and fix $U=8t$, $\Omega=t$, while varying $g$, and $\Omega^{\prime}$, and hole doping $\rho$.

We solved the model using the \gls*{DMRG} method~\cite{DMRG_Steve_white_PRL}, as implemented in the DMRG++ code~\cite{DMRGpp}. Throughout, we primarily focus on \gls*{GS} correlations of the form $\langle \hat{O}_j^{\dagger} \hat{O}_c^{\phantom{\dagger}} \rangle = \bra{\Psi_\mathrm{gs}} \hat{O}_j^{\dagger} \hat{O}_c^{\phantom{\dagger}} \ket{\Psi_\mathrm{gs}}$, where $\ket{\Psi_\mathrm{gs}}$ is the \gls*{GS} wavefunction. (Note the use of the center-site approximation~\cite{white2004real, Nocera2016spectralfunction}, where $c$ is the center site of the chain and $j$ represents sites on the right half of the chain.) In particular, we calculate the spin-spin
\begin{equation}
    C_{\sigma}(r)= \braket{\hat{\bf S}_j \cdot \hat{\bf S}_c},
\end{equation}
density-density
\begin{equation}
C_{\rho}(r)=\braket{\hat{n}_{j}\:\hat{n}_{c}}-\braket{\hat{n}_{j}}\braket{\hat{n}_{c}},
\end{equation}
and bond-bond
\begin{equation}
C_{\mathrm{bond}}(r)=\braket{B_j \: B_c}-\braket{B_j}\braket{B_c}
\end{equation}
correlation functions, where $r=\vert j-c\vert $ is the distance from the center the chain, $B_i = c_{i,\uparrow}^{\dagger}c_{i+1,\uparrow}^{\phantom\dagger}$ is a bond operator, and $\hat{\bf S}_i$ is the usual local spin operator. We calculate the \gls*{SC} correlations from the pair-pair correlation functions~\cite{Banerjee2023groundstate}
\begin{equation}
    C_\text{s(t)}(r)= \braket{\hat{\Delta}_{\text{s(t)},j}^{\dagger}\:\hat{\Delta}^{\phantom\dagger}_{\text{s(t)},c}}, 
\end{equation}
where 
\begin{equation}
\hat{\Delta}_{\text{s(t)},j}^{\dagger}=\frac{1}{\sqrt{2}}[\hat{c}^{\dagger}_{j,\uparrow}\hat{c}^{\dagger}_{j+1,\downarrow}\mp\hat{c}^{\dagger}_{j,\downarrow}\hat{c}^{\dagger}_{j+1,\uparrow}]
\end{equation}
for spin-singlet ($-$) and spin-triplet ($+$) pairing. 

We are also interested in the spin and charge gaps and \gls*{BE} of the model. These quantities are defined as $\Delta_{\mathrm{spin}}(N) = E_0(N,1)-E_0(N,0)$, $\Delta_{\mathrm{charge}}(N) = E_0(N+2,0)+E_0(N-2,0) - 2E_0(N,0)$, and $\Delta_{\mathrm{b}}(N) = E_0(N,0) + E_0(N-2,0) - 2E_0(N-1,1/2)$, respectively, where  $E_0(N,S^z_{\mathrm{tot}})$ is the system's \gls*{GS} energy with $N$ electrons in the $S^z_{\mathrm{tot}}$ magnetization sector. We take the \gls*{BE} in the two-particle limit as a measure of the binding energy in the dilute limit $\Delta_{\mathrm{b}}^{\mathrm{dilute}}\equiv \Delta_{\mathrm{b}}(2,0)$.

Finally, we calculate the dynamical spin structure factor $S(q,\omega)$ using the Krylov space correction vector method~\cite{Nocera2016spectralfunction}. The real space dynamical spin-spin correlation function is given by
\begin{equation}
    S_{jc}(\omega) = -\frac{1}{\pi}\operatorname{Im}\bra{\Psi_\mathrm{gs}}S^z_j
    \frac{1}{\omega -\hat{H}+E_{0}+\mathrm{i}\eta} S^z_c \ket{\Psi_\mathrm{gs}},
    \label{eq:S_ij}
\end{equation}
where the broadening coefficient is set to $\eta=0.1t$. The corresponding dynamical spin structure factor $S(q,\omega)$ is then obtained by Fourier transforming Eq.~\eqref{eq:S_ij}. We also show results for the dynamical charge structure factor in the \gls*{SM}~\cite{Supplement}.

We carried out all \gls*{GS} (excited state) calculations on $L=90$ ($30$) site chains with open boundary conditions. In both cases, we kept $m=500$ \gls*{DMRG} states and a local phonon Hilbert space with $N_{\mathrm{ph}}=10-12$ phonon modes per site, depending on the parameters. These choices maintained a total truncation error below $10^{-7}$.  We have also confirmed that all of our presented results were obtained within the \gls*{SSH} model's physical regime, where the sign of the effective hopping has not been inverted~\cite{Banerjee2023groundstate,Supplement}. \\
\begin{figure}
    \centering
    \includegraphics[width=\columnwidth]{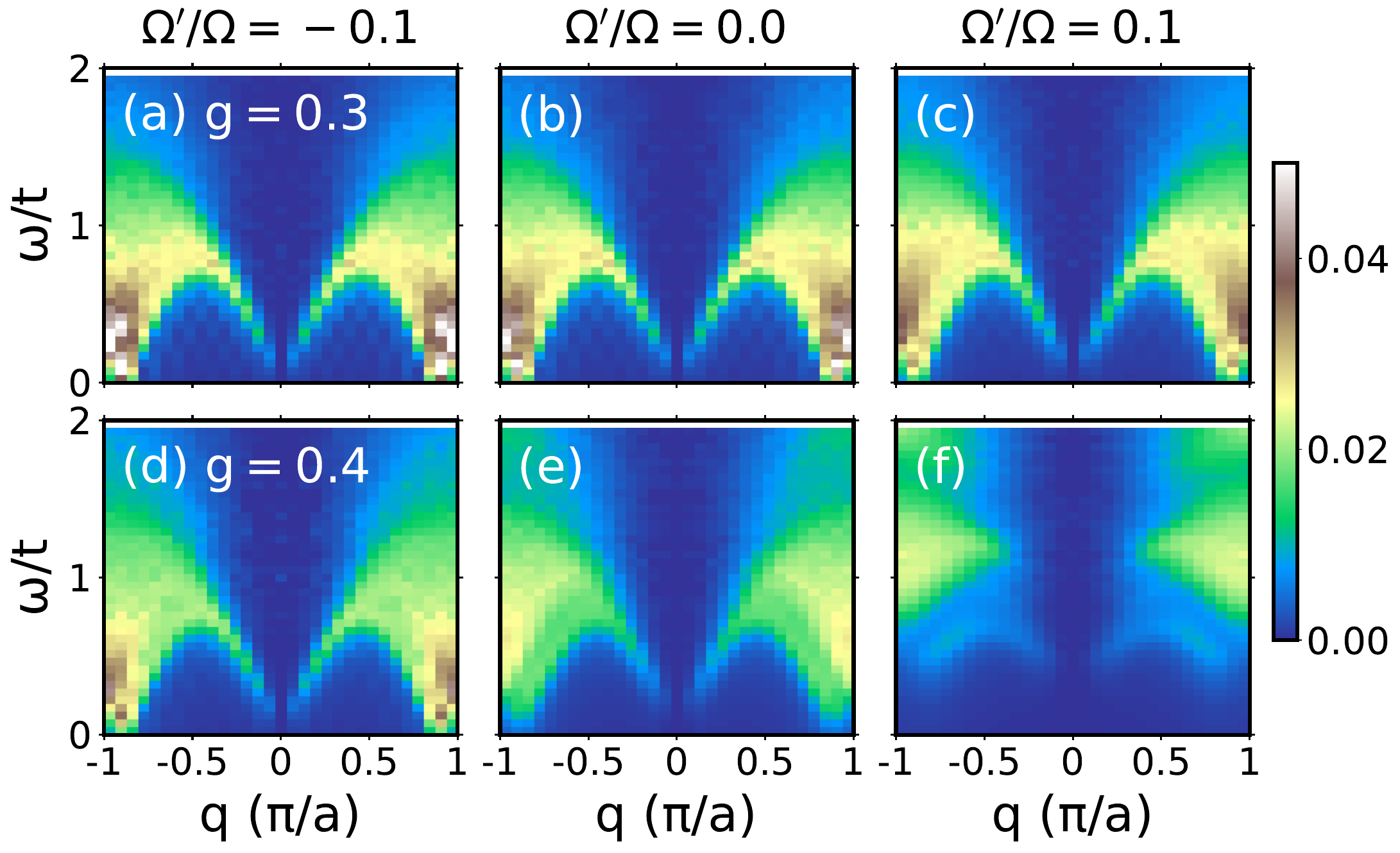}
    \caption{Dynamical spin structure factor $S(q,\omega)$ for the doped HSSH model with dispersive optical phonons calculated for an $L=30$ sites 1D chain with $U=8t$, $\Omega=t$, $\rho = 6.67\%$ with different values of $g$ and $\Omega^{\prime}/\Omega$ as indicated.}
    \label{fig:Sqw_L30_N28}
\end{figure}

%%%%%%%%%%%%%%%%%%%%%%%%%%%%%%%%%%%%%%%%%%%%%%%%%%%%%%%%%%%
% RESULTS
%%%%%%%%%%%%%%%%%%%%%%%%%%%%%%%%%%%%%%%%%%%%%%%%%%%%%%%%%%%
\noindent{\bf Results} --- 
%%%%%%%%%%%%%%%%%%% Figure 1 %%%%%%%%%%%%%%%%%%%%%%%%%%%%%
We first examine the effects of the phonon dispersion on the lattice displacements and electron \glspl*{BE}. Figure~\ref{fig:bare_phonon_dispersion}(b) plots the expectation value of the lattice displacement $\langle \hat{X_i} \rangle$ for $U=8t$, $\Omega=t$, $g=0.4$ and $\rho=6.67\%$. In all cases, the lattice displacements follow a staggered dimerization pattern, modulated by a longer wavelength set by the doping level [recall $2k_\mathrm{F} =\pi(1-\rho)$]. Introducing a dispersion does not change this pattern; however, the amplitude of the displacements increases (decreases) substantially if the phonon energy is decreased (increased) at $q=\pi\approx 2k_\mathrm{F}$. This behavior reflects the reduced cost of creating the displacements associated with the \gls*{BOW}. 

Figure~\ref{fig:bare_phonon_dispersion}(c) compares the \gls*{BE} of the $\rho=6.67\%$ doped \gls*{HSSH} and \gls*{HH} models as a function of $\Omega^{\prime}/\Omega$. Here, negative values indicate that the doped holes bind together. In the \gls*{HSSH} case, $\Delta_\mathrm{b}$ decreases monotonically with increasing $\Omega^{\prime}/\Omega$, which indicates that soft modes near $2k_\mathrm{F}$ result in more effective hole binding. In contrast, $\Delta_\mathrm{b}$ increases with $\Omega^{\prime}/\Omega$ in doped \gls*{HH} model and remains positive for the entire range of $\Omega^{\prime}/\Omega$. Note, we obtain very different behavior in the two-particle (dilute) limit, shown in Fig.~\ref{fig:bare_phonon_dispersion}(d). There, both models predict that $\Delta_{\mathrm{b}}^{\mathrm{dilute}}>0$ over the entire range of $\Omega^{\prime}/\Omega$ values. This result, which stems from the large value of $U$ used here, is consistent with prior work in the dilute limit~\cite{banerjee2025spectral, kovavc2024light}. 

%%%%%%%%%%%%%%%%%%% Figure 2 %%%%%%%%%%%%%%%%%%%%%%%%%%%%%
\begin{figure*}
    \centering
    \includegraphics[width=\textwidth]{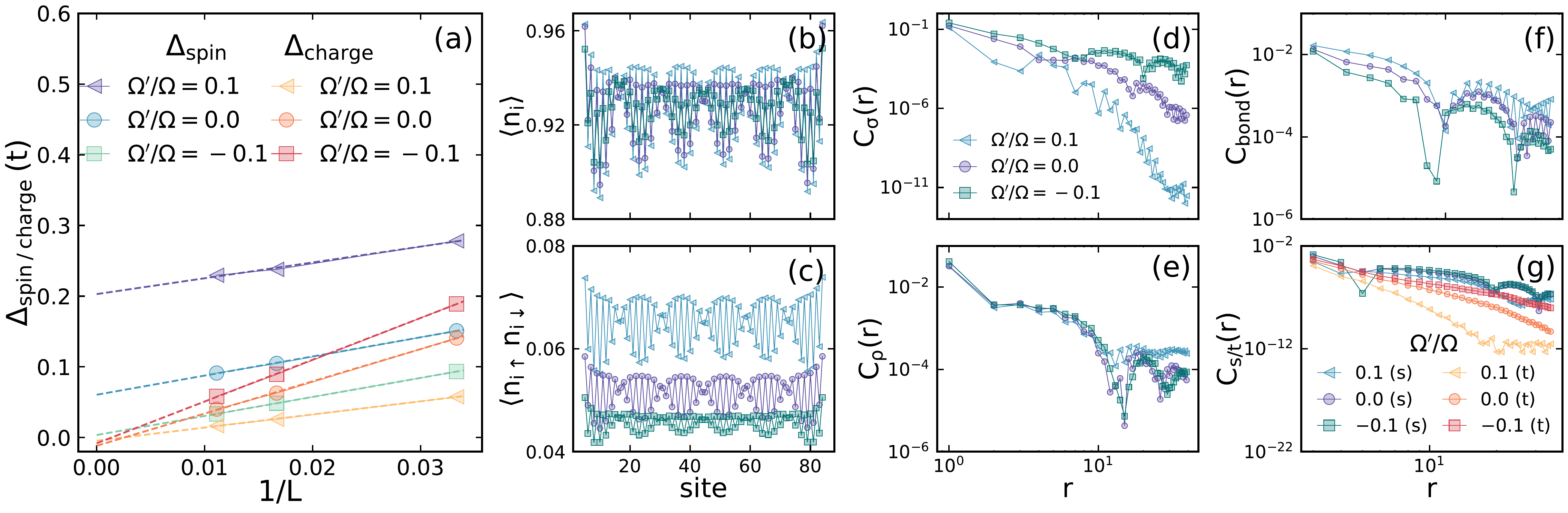}
    \caption{(a) Finite size scaling of the spin $\mathrm{\Delta_{spin}}$ and charge $\mathrm{\Delta_{charge}}$ gaps as a function of cluster size $L$. (b)-(g) Ground state expectation values and correlations functions obtained on an $L=90$ site chain. All results were obtained for $U=8t$, $\Omega=t$, $g=0.4$, $\rho = 6.67\%$, and with varying $\Omega^{\prime}/\Omega$ as indicated in the legend of panels (a), (d), and (g). Panel (b) plots the average electron density, (c) the expectation of double occupancy, (d) spin-spin, (e) density-density, (f) bond-bond, and (g) singlet/triplet correlation functions. The correlation functions in panels (d)-(g) are plotted as a function of $r = j-c$ measured from the chain's center site $c$ and plotted on a log-log scale.}
    \label{fig:GS_correlation_HSSH}
\end{figure*}

Given the value $U = 8t$ considered here, our model has strong \gls*{AFM} correlations that will produce a strong spin response, consistent with measurements on cuprate materials. Motivated by this fact, Fig.~\ref{fig:Sqw_L30_N28} shows $S(q,\omega)$ for $\rho = 6.67\%$ and different values of $g$ and $\Omega^{\prime}/\Omega$, as indicated in each panel. For $g=0.3$ and across all values of $\Omega^{\prime}/\Omega$, the spin structure factor shows the characteristic two-spinon continuum expected for the doped Hubbard model~\cite{li2021particle, karbach1997two}. In this case, doping shifts the nesting momentum from $q=\pi$ to $q=2k_{F}=(1-\rho)\pi$, which results in the gapless excitation at $\pm2k_{F}$ and gapped excitation at $q=\pi$~\cite{kung2017numerically, Parschke2019numerical, Banerjee2023groundstate}. Figs.~\ref{fig:Sqw_L30_N28}(a)-(c) are consistent with these expectations~\cite{voit1995one}; however, for $\Omega^{\prime}/\Omega=0.1$ [Fig.~\ref{fig:Sqw_L30_N28}(c)], the spectral weight at $q=\pi$ shifts slightly higher in energy as compared to the nondispersive case. A similar shift in spectral weight was recently cited as a signature of effective attractive nearest-neighbor interactions in doped Hubbard chains~\cite{shen2024signatures, Li2025doping}. 

\begin{figure}
    \centering
    \includegraphics[width=0.9\columnwidth]{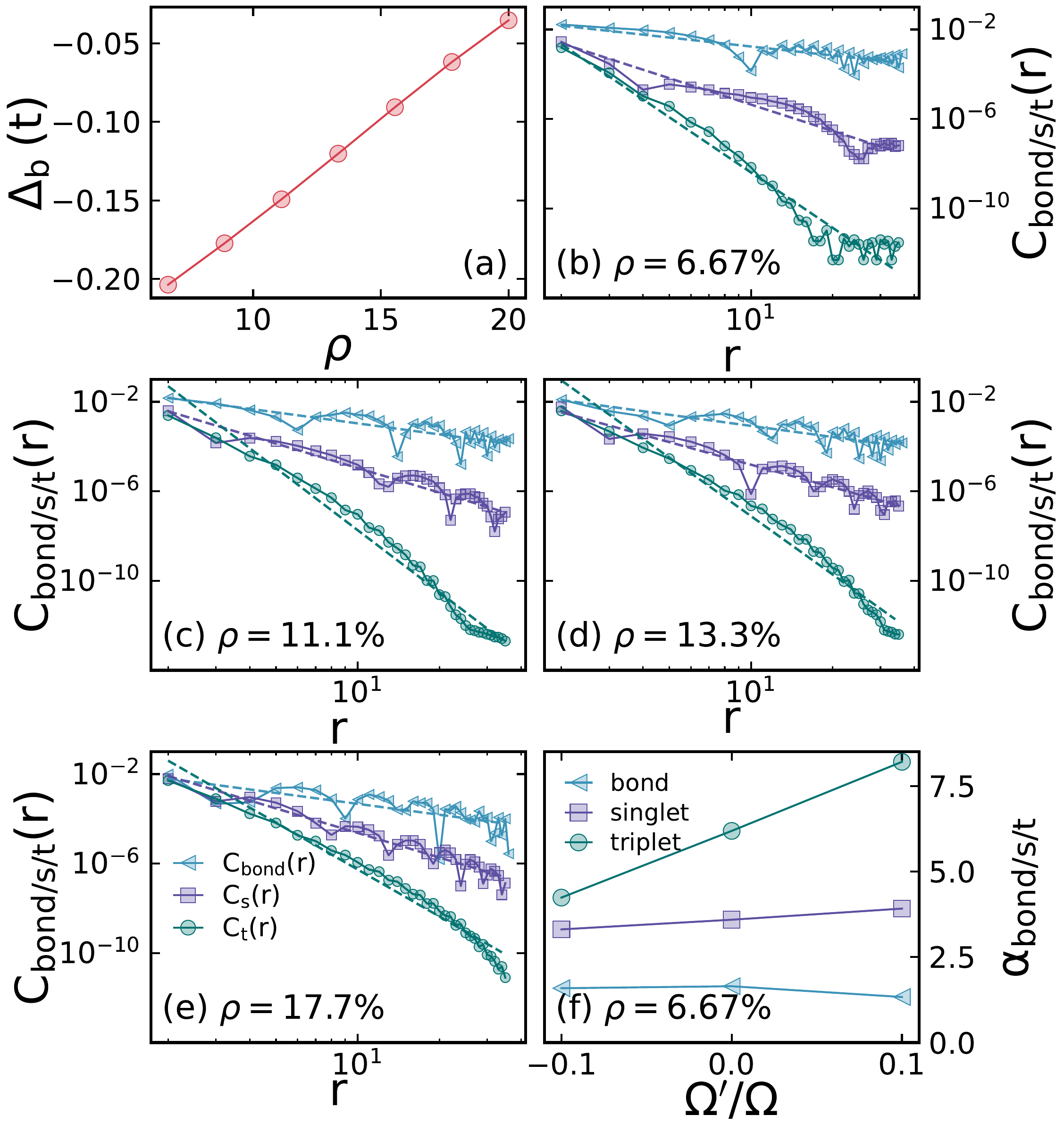}
    \caption{(a) Binding energy of the \gls*{HSSH} model as a function of hole doping $\rho$. (b-e) Bond-bond, singlet, and triplet correlation functions as a function of distance $r$ from the center site, plotted on a log-log scale, for different values of $\rho$ as indicated in each panel. All the results are calculated for an $L=90$ site chain with $U=8t$, $\Omega=t$, $g=0.4$ and $\Omega^{\prime}/\Omega=0.1$. (f) Power-law exponents for bond, singlet, and triplet correlations as a function of $\Omega^{\prime}/\Omega$ for $\rho=6.67\%$.}
    \label{fig:GS_correlation_HSSH_tph01_g04_varydoping}
\end{figure}

Figures~\ref{fig:Sqw_L30_N28}(d)-(f) show $S(q,\omega)$ for an increased $g=0.4$. For $\Omega^{\prime}/\Omega=-0.1,0.0$, the stronger \gls*{eph} coupling shifts spectral weight to the upper part of the two spinon continuum~\cite{Banerjee2023groundstate}. Remarkably, the spin excitations become fully gapped for $\Omega^{\prime}/\Omega=0.1$ [Fig.~\ref{fig:Sqw_L30_N28}(f)]. We attribute this gap to bond singlet formation as the \gls*{BOW} correlations are enhanced, which we will discuss in more detail in Fig.~\ref{fig:GS_correlation_HSSH}. There are also subtle indications of a spin gap in the dispersionless case for $g=0.4$ [Fig.~\ref{fig:Sqw_L30_N28}(e)], which is reflected in the reduced low-energy spectral weight. Note, the charge structure factor $N(q,\omega)$ remains gapless over the entire range of parameters (see \gls*{SM}~\cite{Supplement}), as expected for a doped system.

%%%%%%%%%%%%%%%%%%% Figure 3 %%%%%%%%%%%%%%%%%%%%%%%%%%%%%
Our results indicate that introducing a phonon dispersion can drive strong binding between electrons in the singlet channel. 
It is then natural to wonder whether the system forms a robust \gls*{BOW} or superconducting state in this case. 
To answer this question, Fig.~\ref{fig:GS_correlation_HSSH} shows the evolution of several \gls*{GS} expectation values and correlation functions for fixed $g = 0.4$, $\rho = 6.67\%$ and varying $\Omega^\prime$. Fig.~\ref{fig:GS_correlation_HSSH}(a) plots $\Delta_{\mathrm{spin}}$ and $\Delta_{\mathrm{charge}}$ as a function of the chain length. 
Both quantities have a linear finite-size cluster dependence, which we can extrapolate to the thermodynamic limit  $1/L = 0$, as indicated by the dashed lines. $\Delta_{\mathrm{charge}}$ vanishes for all values of $\Omega^{\prime}/\Omega$ in the thermodynamic limit, consistent with gapless charge excitations, while $\Delta_{\mathrm{spin}}$ remains finite for $\Omega^{\prime}/\Omega=0.0,0.1$. The largest spin gaps occur when the phonon branch has a local minimum near $q=2k_F$ and vanishes for $\Omega^{\prime}/\Omega = -0.1$, consistent with our results for $S(q,\omega)$.

To gain further insight into the origin of the spin gap, Figs.~\ref{fig:GS_correlation_HSSH}(b)-(g) show results for average electron density $\langle n_i \rangle$ and double occupancy $\langle n_{i,\uparrow}n_{i,\downarrow} \rangle$, along with the spin, charge, bond, and \gls*{SC} correlation functions, respectively. Both $\langle n_i\rangle$ and $\langle n_{i\uparrow}n_{i\downarrow}\rangle$ follow the same staggered dimerization pattern seen in Fig.~\ref {fig:bare_phonon_dispersion}b. The amplitude of the oscillations in both quantities increases as $\Omega^{\prime}/\Omega$ is tuned from negative to positive values, together with the average value of the double occupancy. The latter observation indicates that $\Omega^{\prime}/\Omega > 0$ increases the effective attraction between carriers~\cite{banerjee2025spectral,sous2018light}. 

The spin-spin correlation function [Fig.~\ref{fig:GS_correlation_HSSH}(d)] shows a faster decay as a function of $r$ for $\Omega^{\prime}/\Omega=0.1$. This corresponds to the opening of the spin gap, consistent with Fig.~\ref{fig:Sqw_L30_N28}(f) and Fig.~\ref{fig:GS_correlation_HSSH}(a). Whereas the slower decay of $C_{\sigma}(r)$ for $\Omega^{\prime}/\Omega=-0.1$ indicate long-range \gls*{AFM} correlations, consistent with a gapless spin sector. The density-density correlation function [ Fig.~\ref{fig:GS_correlation_HSSH}(e)] shows a rapid algebraic decay for all values of $\Omega^{\prime}/\Omega$, confirming that charge correlations are long ranged and the charge sector is gapless~\cite{Supplement}. The bond-bond correlation function [Fig.~\ref{fig:GS_correlation_HSSH}(f)] shows a slower decay for $\Omega^{\prime}/\Omega=0.1$ compared to other values of $\Omega^{\prime}/\Omega$, indicating that $\Omega^{\prime}>0$ enhances \gls*{BOW} tendencies. Lastly, both singlet and triplet \gls*{SC} correlations [Fig.~\ref{fig:GS_correlation_HSSH}(g)] are weak and decay more rapidly as a function of distance for all values of $\Omega^\prime/\Omega$.  

Taken together, these results show that the $\rho =6.67\%$ doped \gls*{HSSH} model has dominant bond correlation characterized by robust bond singlets, and that soft phonon modes near $q=2k_{F}$ will enhance these correlations. 
Figure~\ref{fig:GS_correlation_HSSH_tph01_g04_varydoping} explores what happens at other doping levels for $\Omega^\prime/\Omega = 0.1$ and $g = 0.4$. (This case has the largest spin gap for $\rho =6.67\%$.) Fig.~\ref{fig:GS_correlation_HSSH_tph01_g04_varydoping}(a) shows \gls*{BE} as a function of $\rho$, where we observe linearly decreasing absolute \gls*{BE} ($|\Delta_\mathrm{b}|$) with progressive doping; however, the \glspl*{BE} remains negative up to the highest doping level we have checked ($\rho=20\%$). Figs.~\ref{fig:GS_correlation_HSSH_tph01_g04_varydoping}(b)-(e) compare the bond-bond (light blue $\triangleleft$), singlet (purple $\square$), and triplet (green $\circ$) correlation functions, respectively, as a function of distance $r$ from the center site. The bond-bond correlation dominates over the \gls*{SC} correlations for all values of $\rho$. The dashed lines in Figs.~\ref{fig:GS_correlation_HSSH_tph01_g04_varydoping}(b)-(e) represent fits of the form $C(r) = Ar^{-\alpha}$. We find that $\alpha_\text{bond} < \alpha_{s/t}$ at all doping levels, indicating again that the bond correlations dominate. For example, for $\rho = 6.67\%$, we obtain $\alpha_\text{bond} = 1.33$, $\alpha_\text{s} = 3.92$, and $\alpha_\text{t} = 8.20$. We show $\alpha_{\text{bond/s/t}}$ values as a function of $\Omega^{\prime}/\Omega$ in Figs.~\ref{fig:GS_correlation_HSSH_tph01_g04_varydoping}(f) for $\rho = 6.67\%$.\\

%%%%%%%%%%%%%%%%%%%%%%%%%%%%%%%%%%%%%%%%%%%%%%%%%%%%%%%%%%%
% DISCUSSION
%%%%%%%%%%%%%%%%%%%%%%%%%%%%%%%%%%%%%%%%%%%%%%%%%%%%%%%%%%%
\noindent{\bf Summary \& Discussion} --- 
We have studied the \gls*{1D} doped \gls*{HSSH} model with dispersive optical phonons with \gls*{DMRG}. Particularly, we have presented results for the \gls*{GS} and dynamical properties of the system with varying \gls*{eph} coupling and finite phonon dispersion in the strongly correlated limit ($U=8t$). We have shown that a positive (negative) phonon dispersion ($\Omega^{\prime}/\Omega$) results in a soft (hard) phonon mode near the \gls*{BOW} ordering vector $q=2k_\mathrm{F}$, which has a direct effect on the system's lattice displacements and enhancement of the bond correlations in the system. Previous studies involving the \gls*{SSH} interaction with Einstein phonons in the dilute limit reveal that \gls*{SSH} phonons can mediate attractive interaction between carriers, leading to small and lightweight bipolarons~\cite{sous2018light,zhang2023bipolaronic,banerjee2025spectral}. Here, we find that even in the light to moderate hole-doped limit, the system's \gls*{BE} decreases with increasing $\Omega^{\prime}/\Omega$, which indicates that strong phonon-mediated attractive interaction persists up to carrier concentrations relevant to synthesized doped cuprate chain compounds~\cite{Chen2021anomalously}. The increased carrier binding occurs in the singlet channel and results in robust \gls*{BOW} correlations and a spin gap in the magnetic excitation spectrum, which can be measured experimentally with \gls*{INS}, \gls*{RIXS}~\cite{banerjee2025identifying, thomas2025theory}, or other techniques. However, for our parameter regime, the charge sector remains gapless~\cite{Supplement} and the \gls*{SC} correlations remain weak. Overall, our results reveal that dispersive optical phonons provide an additional degree of control over binding and spin gap formation for \gls*{1D} correlated systems with \gls*{SSH} interaction. These results also indicate that it may be necessary to go beyond the Einstein phonon approximation when modeling correlated materials. \\

\noindent{\bf Acknowledgments}: 
This work was supported by the National Science Foundation under Grant No. DMR-2401388. A.~N. acknowledges the support of the Canada First Research Excellence Fund. \\

\noindent{\bf Data Availability}: The data supporting this study will be deposited in an online public repository once the final version of the paper is accepted. Until then, the data will be made available upon request. 

\bibliography{references.bib}
\end{document}

\preprint{}
\title{Supplementary Material for ``Enhanced carrier binding and bond correlations in the {H}ubbard-{S}u-{S}chrieffer-{H}eeger model with dispersive optical phonons''}

\author{Debshikha Banerjee\orcidlink{0009-0001-2925-9724}}
\affiliation{Department of Physics and Astronomy, The University of Tennessee, Knoxville, Tennessee 37996, USA}
\affiliation{Institute for Advanced Materials and Manufacturing, University of Tennessee, Knoxville, Tennessee 37996, USA\looseness=-1}

\author{Alberto Nocera\orcidlink{0000-0001-9722-6388}}
\affiliation{Department of Physics Astronomy, University of British Columbia, Vancouver, British Columbia, Canada V6T 1Z1}
\affiliation{Stewart Blusson Quantum Matter Institute, University of British Columbia, Vancouver, British Columbia, Canada
V6T 1Z4}

\author{Steven Johnston\orcidlink{0000-0002-2343-0113}}
\affiliation{Department of Physics and Astronomy, The University of Tennessee, Knoxville, Tennessee 37996, USA}
\affiliation{Institute for Advanced Materials and Manufacturing, University of Tennessee, Knoxville, Tennessee 37996, USA\looseness=-1}

\date{\today}% It is always \today, today,
             %  but any date may be explicitly specified

\maketitle
\color{black}

\section{Critical electron-phonon coupling for the doped HSSH model with dispersive optical phonons}

\begin{figure}[h]
    \centering
    \includegraphics[width=\textwidth]{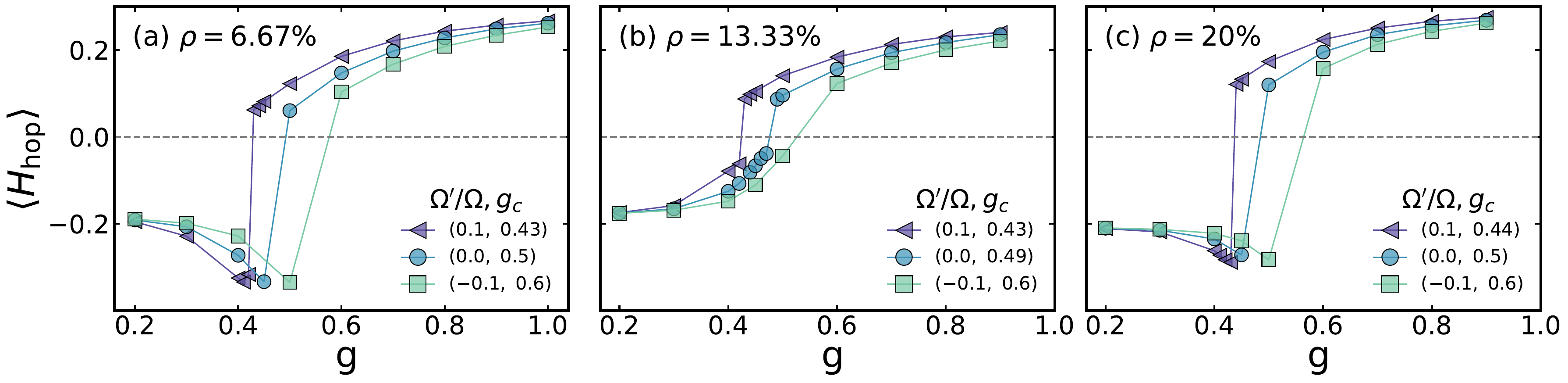}
    \caption{Ground state expectation values of single-particle hopping $\langle H_{\mathrm{hop}} \rangle$ for the \gls*{HSSH} model with dispersive optical phonons, calculated with DMRG for an $L=30$ site chain with $U=8t$, $\Omega=t$ and varying $\Omega^{\prime}/\Omega$ for different values of $\rho$ as indicated in the panels. The expectation values were calculated for hopping between the center site and its neighboring site as a function of $g$. The expectation values become positive when the effective hopping changes sign, as discussed in Ref.~\cite{Banerjee2023groundstate}. For $g > g_c$, the linear \gls*{SSH} model is no longer valid, and the model becomes unphysical.} 
    \label{fig: critical_coupling}
\end{figure}

The modulation of the hopping integral in the \gls*{SSH} model is derived from a linear approximation for the distance dependence of the overlap integral. We can, therefore, define an effective hopping as  
$t_{\mathrm{eff}} \approx -t + g \langle X_i - X_{i+1} \rangle$~\cite{Banerjee2023groundstate}. For a strong \gls*{eph} coupling, the average lattice displacements tend to increase and can become large enough to change the sign of $t_{\mathrm{eff}}$. When this occurs, the system is unstable toward dimerization and the linear \gls*{SSH} model becomes unphysical~\cite{Nocera2021bipolaron, Banerjee2023groundstate}. The critical coupling $g_\mathrm{c}$ for such sign inversions can be determined by monitoring the average value of the single-particle hopping at the chain's center~\cite{Banerjee2023groundstate}. To this end, Fig.~\ref{fig: critical_coupling} plots $\langle H_{\mathrm{hop}}\rangle=\langle \hat{c}_{c-1,\uparrow}^{\dagger}\hat{c}_{c,\uparrow}^{\phantom{\dagger}}\rangle$ as a function of \gls*{eph} coupling $g$ for an $L=30$ site chain with $U=8t$, $\Omega=t$, and for varying $\Omega^{\prime}/\Omega$ and $\rho$ as indicated in the panels. Increasing $\Omega^{\prime}/\Omega$ shifts the critical coupling $g_\mathrm{c}$ to lower values. This behavior is easily understood by recognizing that placing a soft phonon mode at $q \approx 2k_\mathrm{F}$ reduces the lattice potential energy required to form large dimer displacements. 

\newpage
\section{Power law fitting exponents for different correlation functions}
\begin{figure}[h]
    \centering
    \includegraphics[width=0.7\textwidth]{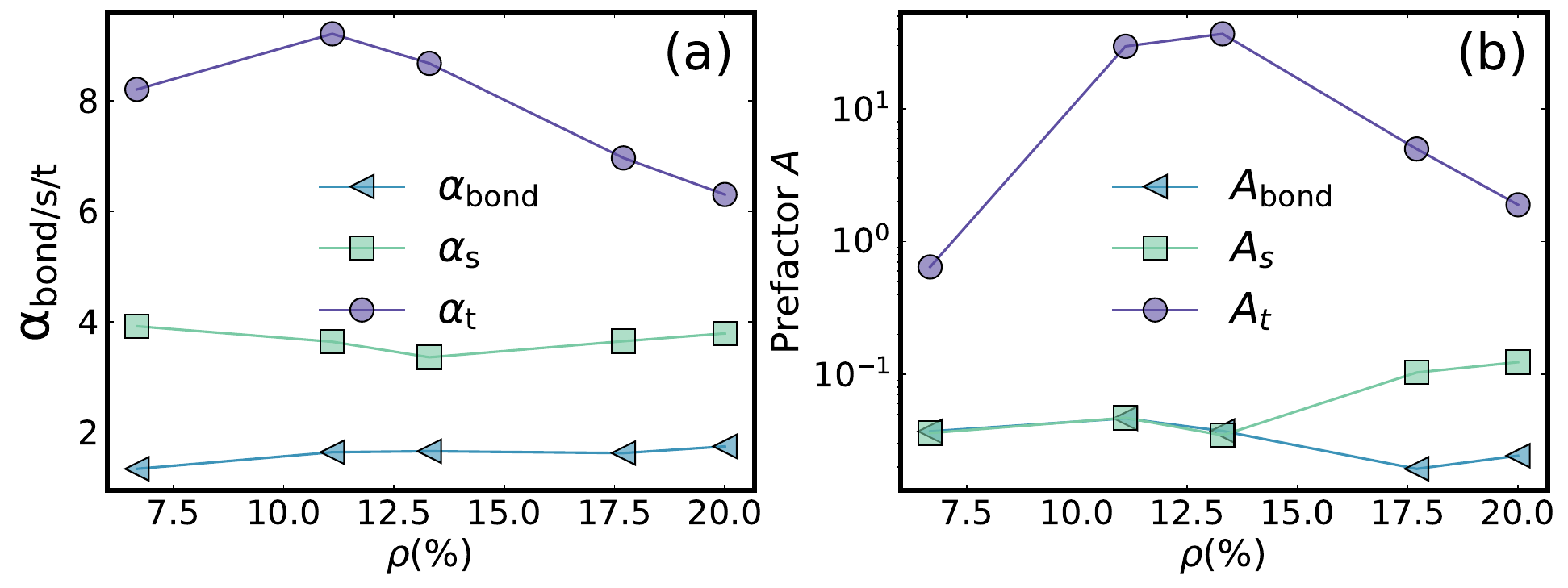}
    \caption{Fitted power law (a) exponents $\alpha$ and (b) prefactors $A$ for bond-bond, singlet, and triplet correlations plotted as a function of doping $\rho$. These parameters were obtained by fitting the corresponding correlation functions calculated on an $L=90$ site chain with $U=8t$, $g=0.4$, $\Omega=t$, and $\Omega^{\prime}/\Omega=0.1$.}
    \label{fig:exponent_doping}
\end{figure}

Figure~\ref{fig:exponent_doping} shows the power law  exponents ($\alpha_{\text{bond/s/t}}$) and prefactors ($A_{\text{bond/s/t}}$) obtained from fitting the correlation functions with the form $C(r) = Ar^{-\alpha}$ as a function of doping of $\rho$ in the region $2 \le r \le 35$, as shown in Fig.~4 of the main text. In this case, all fits were performed on data calculated on an $L=90$ site chain with $U=8t$, $g=0.4$, $\Omega=t$, and $\Omega^{\prime}/\Omega=0.1$. 
Figure \ref{fig:exponent_doping} shows that even though $A_{\text{s}}=A_{\text{bond}}$ for up to $\rho=13.3\%$, $\alpha_{\text{bond}} < \alpha_{\text{s}}$ for all values of $\rho$, indicating that the superconducting correlations decay faster as a function of distance. Conversely $\alpha_{\text{t}}$ shows nonmonotonic change with increasing $\rho$ but with much higher values than both $\alpha_{\text{bond}}$ and $\alpha_{\text{s}}$. These results provide further evidence of the enhancement of bond correlations in the model.

\newpage
\section{Dynamical charge structure factors}
\begin{figure}[h]
    \centering
    \includegraphics[width=0.7\textwidth]{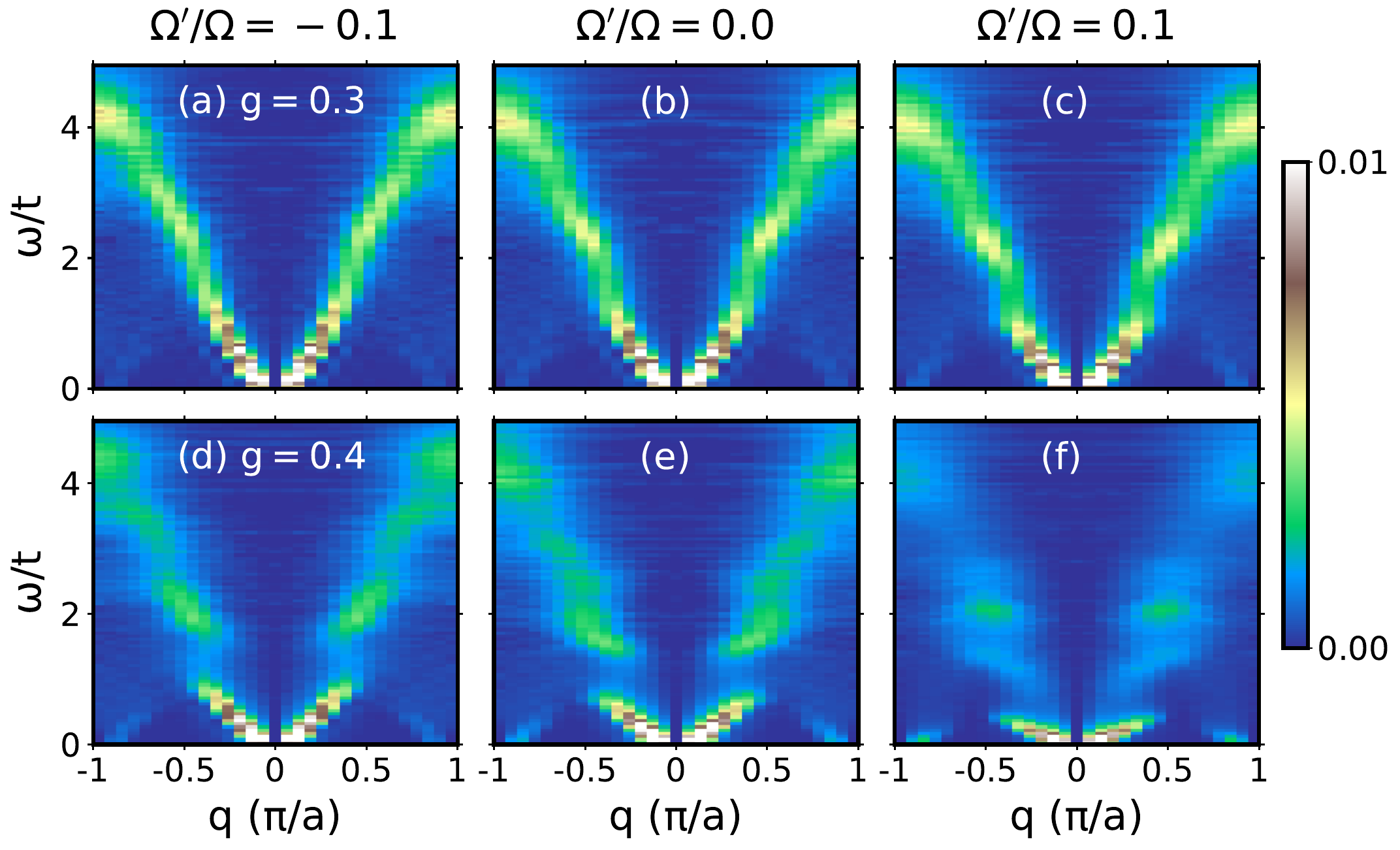}
    \caption{Dynamical charge structure factor $N(q,\omega)$ for the doped HSSH model with dispersive optical phonons calculated for an $L=30$ sites 1D chain with $U=8t$, $\Omega=t$, $\rho=6.67\%$, and with different values of $g$ and $\Omega^{\prime}/\Omega$ as indicated. Panels (a-c) show results for $g=0.3$ and (d-f) $g=0.4$.} 
    \label{fig:L30_N28_Nqw}
\end{figure}

Figure~\ref{fig:L30_N28_Nqw} shows results for the dynamical charge structure factor $N(q,\omega)$ of the doped \gls*{HSSH} model with a dispersive optical phonon branch. The results are calculated for an $L=30$ site chain with $U=8t$, $\Omega=t$, $\rho=6.67\%$, and with different values of $g$ and $\Omega^{\prime}/\Omega$ as indicated in the figure. Similar to $S(q,\omega)$, we calculate $N(q,\omega)$ with \gls*{DMRG} using the Krylov space correction vector method~\cite{Nocera2016spectralfunction}. The real space dynamical charge correlation function is given by
\begin{equation}
    N_{cj}(\omega) = -\frac{1}{\pi}\operatorname{Im}\bra{\Psi_\mathrm{gs}}\Tilde{n}_j
    \frac{1}{\omega -\hat{H}+E_{0}+\mathrm{i}\eta} \Tilde{n}_c \ket{\Psi_\mathrm{gs}}.
    \label{eq:N_ij}
\end{equation}
Here, $c$ is the center site of the chain, $\ket{\Psi_{\mathrm{gs}}}$ is the ground state of the system with energy $E_0$, and $\Tilde{n}_j = (\hat{n}_j-\langle \hat{n}_j \rangle)$.

For $g=0.3$, $N(q,\omega)$ shows a continuum of charge excitations extending up to the free electron bandwidth $4t$, with maximum spectral weight appearing near the lower energy part of the spectrum. We also observe kink-like features appearing at multiples of phonon energy. 
As discussed in Ref.~\cite{Banerjee2023groundstate}, these features reflect modifications of the charge continuum due to the usual dispersion kinks observed in \gls*{eph} coupled systems. Increasing coupling to $g=0.4$ decreases spectral weight at higher energy and enhances the kink features across all values of $\Omega^{\prime}/\Omega$. Importantly, we observe no charge gap in the system for all values of $g$ and $\Omega^{\prime}/\Omega$, indicating that charge excitations remain gapless. 

\newpage
\section{Dynamical spin structure factors}
\begin{figure}[h]
    \centering
    \includegraphics[width=\textwidth]{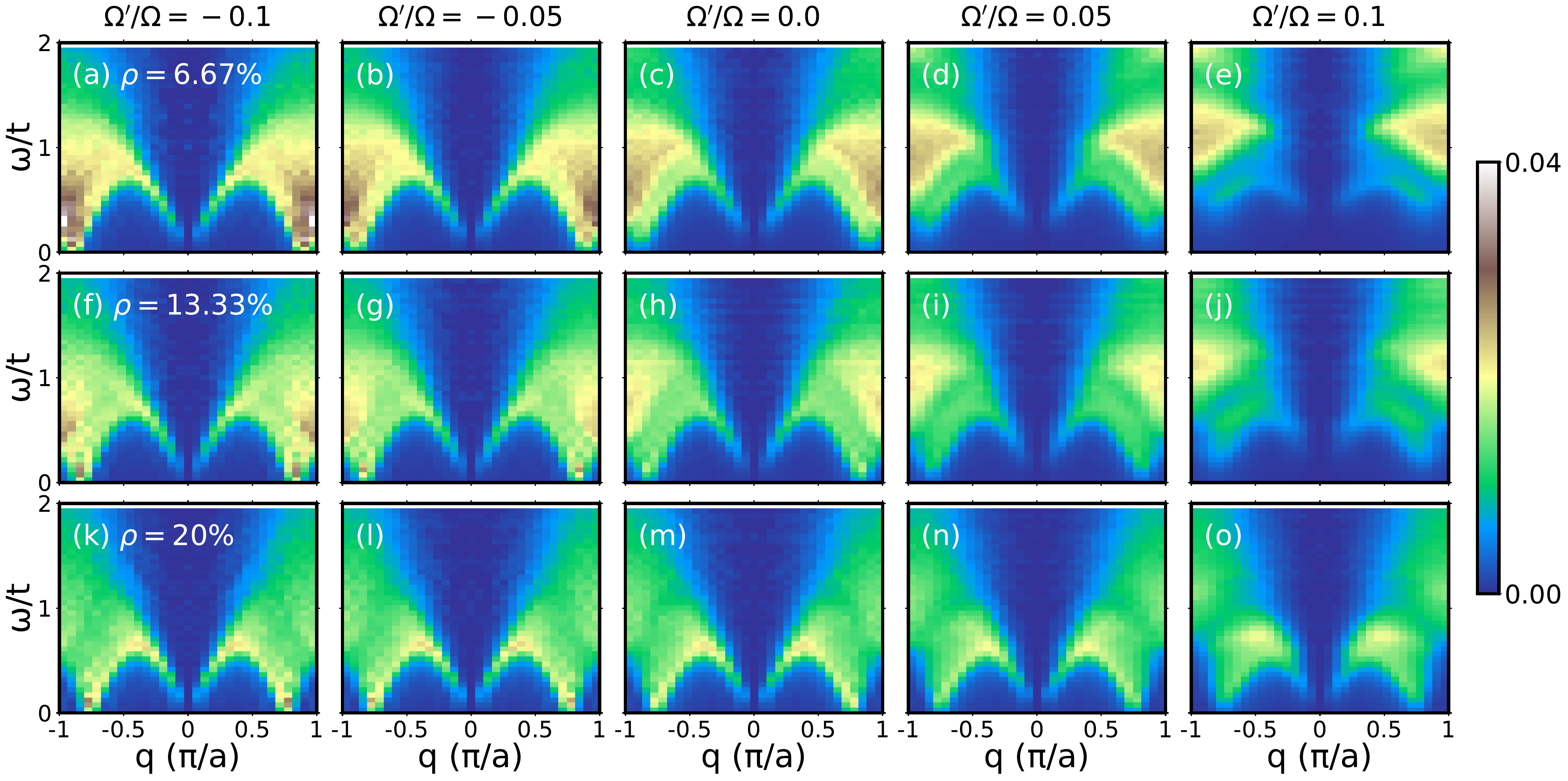}
    \caption{Dynamical spin structure factor $S(q,\omega)$ for the doped HSSH model with dispersive optical phonons calculated for an $L=30$ sites chain with $U=8t$, $\Omega=t$, $g=0.4$, and with different values of $\rho$ and $\Omega^{\prime}/\Omega$ as indicated. Results are shown for (a)-(e) $\rho=6.67\%$, (f)-(j) $\rho=13.33\%$, and (k)-(o) $\rho=20\%$ hole doping values.}
    \label{fig:all_doping_Sqw_HSSH}
\end{figure}

Figure~\ref{fig:all_doping_Sqw_HSSH} shows additional results for the dynamical spin structure factor $S(q,\omega)$ of the doped \gls*{HSSH} model with dispersive optical phonons. The results are calculated for an $L=30$ site chain with $U=8t$, $\Omega=t$, $g=0.4$ and with different values of $\Omega^{\prime}/\Omega$ and $\rho$ values.
In this case, we observe prominent spin gaps for $\Omega^{\prime}/\Omega=0.05,0.1$ for $\rho=6.67\%$. However, the spin gap diminishes with increasing $\rho$, reflecting the decrease in binding seen in Fig.4 in the main text. For example, for $\rho=20\%$  and $\Omega^{\prime}/\Omega=0.1$, there remains a subtle signature of the spin gap in the spectrum.

\bibliography{references}